# ROLE OF DIGITAL PLATFORMS IN ENTREPRENEURIAL PROCESSES: A RESOURCE ENABLING PERSPECTIVE OF STARTUPS IN PAKISTAN

Hareem Nassar, NUST Business School, Pakistan, hareem.msie19nbs@student.nust.edu.pk

Fareesa Malik, NUST Business School, Pakistan, fareesa.malik@nbs.nust.edu.pk

**Abstract:** This article aims to explore the role of digital platforms as external enablers in entrepreneurial processes. The recent infusion of digital platforms into different aspects of innovation and entrepreneurship has supported digital entrepreneurship; however, the altered entrepreneurial processes are yet to be explored. This study focuses on digital platform-based startups of Pakistan and draws on entrepreneurial bricolage theory to understand the enabling external resources. We followed multiple qualitative case studies approach and collected data through semi-structured interviews from two startups operating solely on digital platforms, 1) XYLEXA and 2) Toycycle. The findings show that entrepreneurial process is a continuous process. Digital platforms have made entrepreneurial processes less bounded i.e. the products and services keep on evolving even after they have been endorsed to the end user. Moreover, platform-based startups having limited resources can move through the entire entrepreneurial process by combining available resources efficiently and effectively.

## 1. INTRODUCTION

The recent infusion of digital platforms into different facets of innovation and entrepreneurship has transformed the nature of uncertainty inherent in the entrepreneurial processes along with the ways of dealing with such uncertainty [13, 19, 23, 33, 36]. This has opened up some essential research directions, at the intersection of digital platforms and entrepreneurship i.e. digital entrepreneurship, which considers digital platforms and their distinctive features in influencing entrepreneurial pursuits [24]. Digital platforms have not only shaped the entrepreneurial processes (opportunity generation, opportunity development and opportunity exploitation) but have also brought changes in innovation, competences, control, financing, institutions and ecosystems [34]. Digital entrepreneurship includes transforming existing businesses or new ventures with the help of digital technologies. It is viewed as a vital pillar for development in the digital economy [34]. Digitalization has rendered entrepreneurial processes less bounded i.e. there has been a shift from discrete and steady boundaries to highly porous and fluid boundaries which enables the products and services to continuously evolve even after they have been introduced in the market [24]. Digitalization of entrepreneurial processes has also helped in breaking down the boundaries between various phases along with bringing greater levels of unpredictability and nonlinearity into how they fold [24].

Businesses operating on digital platforms are quite different from the traditional businesses in terms of building trust, governance, resources and entrepreneurial processes. The study explores this through the theoretical lens of entrepreneurial bricolage which explains how entrepreneurship can be done through minimal resources [2, 14, 16, 27, 42]. Entrepreneurial bricolage can be a feasible path for platform-based startups or SMEs, having limited resources, to help in facilitation of entrepreneurial processes.

Even though, a considerable number of entrepreneurs and businesses are using digital platforms to tap opportunities, research is still quite limited in this context [34]. In Pakistan, digital platforms have originated in recent years. Since many digital platform-based startups have started operating in Pakistan and various SMEs are also shifting their businesses on platforms, it becomes an important research area to explore. The study focuses on the following research question:

"How do digital platforms act as external enablers in entrepreneurial processes?"

To find answers to the above research question, we conducted an interpretive and qualitative study. We selected two platform-based startups operational in Pakistan: 1) Toycycle and, 2) XYLEXA for comparative case studies. The article explores the role of digital platforms as facilitators of various resources for startups. It also highlights the resource challenges that platform-based startups face in the execution and implementation of their ideas.

## 2. LITERATURE REVIEW

### 2.1 Digital Platforms

Digital platforms are characterized as a sociotechnical grouping which includes the technical elements of software and hardware as well as the organizational processes and principles [6, 28]. They are a shared and common set of services and architecture that provide help in hosting complementary offerings [24]. Digital platforms and related ecosystems are often marked by the role of a single firm, the platform leader, in creating the modular platform and in generating both value creation and value appropriation [24, 25]. Digital platforms serve to be infrastructure, marketplace and ecosystems at the same time. For instance, Facebook and Google are digital platforms which provide social media and search but at the same time, they also serve to be the platforms on which other platforms can be built [18]. They have flourished as engines of innovation so that other firms can build complementary products and services in ecosystems [36]. Although the concern regarding the governance of digital platforms is a prevailing issue [28], digital platforms still have the potential of disrupting traditional business models, organizations and all other forms of value creation and capture. They filter and customize information, which are shared by many companies within same or different industry, and also take the form of business community platforms, which are personalized for usage by all the members of a particular business community [18, 21].

Modular systems are leading to the development of platform architectures [36, 39]. In digital platform sites, there is greater interdependence between entrepreneurial firms that launch specific modules and platform firms for whom the modules are launched. Platform firms spend significant amount of resources to attract third-party developers to their platforms to get support from them and build a higher installed base which incentivizes entrepreneurs to introduce more complementary modules [1, 11, 39].

Digital platforms having large user base is more valued by entrepreneurs as they have the largest potential market for their complementary products [39, 41]. They create indirect network effects [7, 22, 39, 40] which serves as the basis of competition in digital platform settings. The choice of an entrepreneur to support the platform is greatly influenced by the network effects for the platform. The presence of network effects and installed base advantages are vital elements of success in platform industries, leading to many new platforms and competitors [39, 41].

### 2.2 Entrepreneurial Processes

The entrepreneurship process is an activity which processes the opportunities. It goes through the process of opportunity generation (creation and discovery), opportunity development and opportunity exploitation with the objective to transform an opportunity into a viable venture and thus, achieve success [13, 33].

Previous studies on innovation and entrepreneurship as well as the present theories on product life cycle, architectural innovation and product development process have assumed constant and discrete boundaries for ideas relating to new products and services that underlie an entrepreneurial opportunity [8, 24, 35]. However, infusion of digital technologies has made these boundaries more permeable as the scope, attributes and importance of product or service keep evolving even after the idea has been endorsed. For instance, Tesla has introduced various new functions and features in its cars even after they have been endorsed to the market, simply by modifying digital artifacts or components. With digital technologies and platforms, entrepreneurial processes have also become less bounded as they allow ideas and business models to be formed, endorsed, amended and restructured rapidly, e.g. 3D printing [24]. The scalability of digital platforms (e.g. cloud computing and mobile networking) also causes variations in entrepreneurial activities [24]. For instance, Airbnb started with its primary attention on providing hotel space for various meetings and events. Later, it catered to the demand for affordable accommodation which the hotels were unable to meet thus, rapidly scaling up its services enabled by cloud computing services. Thus, digital technologies infuse a greater level of fluidity and variability into entrepreneurial processes. These changes in entrepreneurial processes enabled by digital platforms lead to change in behaviors and actions of entrepreneurs in the digital arena [24].

With traditional models and frameworks on entrepreneurship assuming fixed and stable boundaries for an entrepreneurial opportunity, a more evolving stream in entrepreneurship research presents alternate views regarding opportunity creation and enactment that reflects fluid boundaries for entrepreneurial processes. For instance, the perspective of 'opportunity creation' is of the view that opportunities are emergent and the entire creation process is evolutionary [14, 24]. Likewise, the 'effectuation' perspective suggests that the entrepreneur continuously re-evaluates all the available means and shape the offering [24, 30]. The 'narrative' perspective makes sense of the meaning associated with entrepreneurial opportunities [15, 24]. All these perspectives indicate that there are fluid boundaries with respect to entrepreneurial processes.

Thus, it is concluded that alternative concepts and theories are necessary for integrating new ways of evaluation of entrepreneurial success and inform on all those factors that are linked with progression of entrepreneurial processes. Digital platforms play a major role in shaping such liminal entrepreneurial processes [24].

### 2.3 Entrepreneurial Bricolage Theory

The "theory of entrepreneurial bricolage" allows entrepreneurs to endure or even establish strong and growing firms in spite of scarce resources [37, 38]. It allows entrepreneurs to build available resources in an innovative manner into new products or services rather than merely accepting their current potential [2, 43]. The theory of entrepreneurial bricolage has three important features. (1) assessing whether an effective outcome can be generated from what is currently available. (2) combining and orchestrating resources in an innovative manner for new applications rather than only using them for their originally intended purposes. (3) using available resources rather than looking for new resources [43]. Startups are generally very resilient, flexible and creative [4, 12, 17] but their limited network and scarce resources pose to be a great challenge for them [10]. With the help of this theory, startups are able to discover many new prospects by overcoming difficulties in resource acquisition [16, 27, 43]. Entrepreneurial bricolage theory also complements with the Resource based view (RBV) and the institutional view. This is because RBV is not much readily applicable in the context of startups as it is quite difficult for startup to acquire unique resources in undeveloped market. Thus, this theory asserts that startups can take benefit from existing under-utilized resources by combining them in unique ways [3, 43]. Moreover, firm's operations being in conformity with traditional values and beliefs create suboptimal resource choices which prevent firms from pursuing economically viable options [5, 26, 32, 43]. Thus, the theory states that startups should cross the traditional boundaries to create new products and services [9, 43].

Based on the nature of resources, entrepreneurial bricolage has been classified into three types: First, the input bricolage which combines physical (materials) and human (labor and skillset) resources in an innovative manner and apply them to new problems and opportunities. It leverages low cost labor for various entrepreneurial activities by making full use of available resources. Input bricolage increases operational efficiency when startups have very limited financial resources or have to react instantaneously to the demands of their customers [43]. Input bricolage also helps platform-based startups in recombining available resources and improving sales performance by broadening the distribution channels, providing infinite shelf space and targeting new audience [14, 43]. Second, the market bricolage transforms existing network of entrepreneurs (customers, friends, suppliers and competitors) to create new customers from that market in which rivals operate. In platform businesses, many customers begin as or become friends. Suppliers become customers and vice versa. Such shifts and expansion of roles deepens the understanding of customer needs and receive feedback from them. Market bricolage enables digital platform-based startups to broaden their product and service combinations at low cost through economies of scale as well as develop trust among business partners [2, 14, 43]. Third, the institutional bricolage incorporates innovative procedures and practices resulting in an institutional transformation. It involves socially reconstructing the available resources and combining them in ways which sets up new institutions. Institutional bricolage is essential for platform-based startups in breaking the inertia of routines [2, 9, 43].

## 3. METHODOLOGY

We adopted a qualitative multiple case study approach to deeply explore the role of digital platforms in entrepreneurial processes of platform-based startups in developing countries like Pakistan as it is a context-based research. Considering the early stage of establishment of the startups, multiple case study approach was more suitable to develop an in-depth understanding of the phenomena than a single case could provide and to explore the answers of 'how' questions for theory building. The evidence created from a multiple case study is strong and reliable and similarities and contrasts can be made. Moreover, this approach creates a more convincing theory when the suggestions are intensely grounded in several empirical evidence, thus allowing for a wider exploration of our research question and theoretical evolution [31]. We selected two platform-based startups operating in Pakistan, 'Toycycle' and 'XYLEXA', for data collection. Toycycle is an online platform for the buying and selling of preowned items including baby gear (strollers, high chairs, bouncers and carriers), clothes and toys (games, puzzles, electronic toys and wooden toys). Whereas, XYLEXA is an online platform for provision of diagnostic services to caregivers using AI and image processing techniques. The platform serves as a decision support system for radiologists by providing medical image diagnosis and disease and is also involved in R&D for timely diagnosis of cancer.

The primary data was collected through informal chats and semi-structured interviews with founders, co-founders and employees of both startups. Altogether five semi-structured interviews were conducted from founders (2 interviews – both males), co-founders (2 interviews – 1 male and 1 female) and employees (1 interview - male) of both startups. All interviews were conducted in English language which were later transcribed. The transcriptions were read multiple times for thematic analysis. The concepts of ICT and entrepreneurial bricolage theory helped us in making sense of the data. The themes were finalized after extensive discussion within the project team.

## 4. FINDINGS

This section summarizes the research findings in three themes to explain the role of digital platforms as resources enabler in entrepreneurial processes.

### 4.1. Input Bricolage: Combining Internal and External Resources

Building a platform-based startup is challenging in Pakistan. The digital economy of Pakistan is still in the developing phase which leads to technological and acceptance issues. Apart from this, startups also face the issue of resource scarcity due to limited resources. But by fully utilizing some available resources, they move through the entire entrepreneurial process.

When the startups were at the opportunity generation stage i.e. opportunity discovery and creation, they made full use of skills, experience and knowledge base for selecting market and customer problems. As explained by one respondent:

*'Market research was very critical in the starting as we had to get it right to let our customers buy from us. However, I have been in healthcare industry for about 16 years, so I am already familiar with the market and I am bringing in those customer voices and concerns.' (R 3, XYLEXA Co-founder)*

Due to their newness and smallness, it was very expensive for them to seek and acquire resources from their stakeholders. But with the startups operating on third-party digital platforms, friends and family resources came in handy to lower the operational costs by bringing in supportive infrastructure and less expensive labor.

*'We brought in consultants who were relevant in areas of Artificial Intelligence (AI) and Machine Learning to come in and help the team at different stages where they get stuck. They were expensive resources but we were able to negotiate a very good package with them so that they can provide guidance to the team.' (R 2, XYLEXA Co-founder)*

Moving towards opportunity development and exploitation, startups also faced financial and technological resource challenges. They had developed a viable prototype but lacked resources to turn it into a tangible product.

*'We raised two pre-seed rounds for this challenge. As a result, now we are close to our break-even.' (R 4, Toycycle Founder).*

Another respondent also articulated this:

*'Dealing with Machine Learning is very resource intensive. Also, we are using Multiple Languages which is an IBM cloud based application. It is quite expensive but since we have got credits from IBM, it is advantageous for us. We are also stuck with paying license fees.' (R 2, XYLEXA Co-founder)*

Recombining available resources in an effective manner also helped in improving sales performance. Having operations entirely on digital platforms broadened the distribution channels and provided infinite shelf space for new products and services and targeting new audience.

*'As we had compact houses, we were facing the issue of disposing of extra baby gear in an eco-friendly manner. So we started off with the platform for which we allowed people to subscribe and swap things free of cost.' (R 4, Toycycle Founder)*

### 4.2. Market Bricolage: Creating New Customers and Building Trust

As the startups moved from opportunity generation to opportunity development and exploitation process, they felt the need to create new markets and new customers for which they transformed the existing network of entrepreneurs i.e. customers, friends, suppliers and competitors. Customers became friends, suppliers became customers and vice versa.

*'We did partnerships with relevant suppliers and startups to enhance our customer base.' (R 5, Toycycle Employee)*

Aiming for a large user base was an issue as advancement in digital technologies in Pakistan is still at a nascent stage, either because of lack of adequate technological knowledge or because of high costs.

*'Not a single hospital in Pakistan is using CAD system. It is not that they do not know about it. It is just that they cannot afford it. It can cost you as much as $300,000 plus the maintenance charges. So we had to make it very comparative in terms of price point and entry where they do not have to invest any dollars in capital investment, there is no expense and they only have to pay for what they use. Wherever we gave our commercial proposals, nobody said that it is too expensive. They said that it is very reasonable. So we are creating a new market by making sure that we have enough customers who would be our referential customers and that will help us grow and expand our system in the hospitals in Pakistan and globally.' (R 2, XYLEXA Co-founder)*

Through digital platforms, it was easier to shape the offerings i.e. the products and services even after they had been introduced to the end users simply by modifying digital artifacts or components. As customer requirements change with time, entrepreneurship process becomes ongoing and continuous. This also helped in retaining customers and attracting new ones.

*'We generally go through a product management process. We have built a base product; we have requirements coming in from customers whether they are related to user experience or new functionality. We go through a standard release process. We prioritize them and implement them.' (R 2, XYLEXA Co-founder).*

Another respondent also articulated this:

*'We keep our eyes on whether our old product design is up to date and working properly, if we can improve our conversion and customer acquisition. We already have two defined KPIs i.e. the amount of stuff that we pick up, which is also our inventory that we need to sell. The other is sales that we generate. In order to increase sales, we incentivized our check-out page like offering discount to boost our sales because some customers used to add stuff to the cart and left instead of purchasing the items. So this was a way to bring back such customers as well.' (R4, Toycycle Founder)*

Emerging startups also faced low level trust and governance issues as well as less committed relationships between the business partners at the opportunity development and exploitation stage.

*'At this stage, it was very important to build trust. Since I had already been in the industry for 16 years and selling to same hospitals so our customers were all referral customers. This helped us in building trust among them comparatively easily.' (R 3, XYLEXA Co-founder)*

To build an even better level of trust with the customers and business partners, products and services went through an entire process of trial and error and feedback was received from customers. This is how the startups went from concept to proof of concept. Minimum Viable Product (MVP) was released in the market. The needs and concerns of the customers were received and now the beta product is incorporating those customer feedbacks.

### 4.3. Institutional Bricolage: Adoption of Innovative Approaches to Bring an Institutional Transformation

Startups adopted some innovative principles, rules and practices that were away from the traditional ones, to socially reconstruct the resources at hand. This helped them to break the inertia of routines. Various innovative approaches were implemented to employ available resources. Individualized and customized services were made available to the end users and open source was used to build components and add new functionalities. Labor was one of the most important resources so every effort was made to retain them to transform the institution (startup). As one respondent articulated:

*'In order to retain our employees, we offered them trading, a good working environment, the interesting product that we were making and the cutting-edge technology that we are using. We took care of our employees and gave them a chance to learn and grow and not make empty promises to them. We are sitting in the same office, working on the product and funding. So they have seen the progress themselves and they are committed to what we are doing.' (R 2, XYLEXA Co-founder)*

These platform-based startups have become pioneers in their area by introducing novel products in the market. As a result, the perception of people has changed to some extent. They are adopting new technology based products. The startups have also experienced first-mover advantage.

### 5.    DISCUSSION AND CONCLUSION

The study seeks to make contribution to research and practice in digital entrepreneurship. It offers novel insights into digital entrepreneurship literature by exploring role of digital platforms as facilitators of various resources for startups.

The answer to the research question – "How digital platforms act as external enablers in entrepreneurial processes?" has been examined with the help of entrepreneurial bricolage theory. Digital platforms serve to be infrastructure, marketplace and ecosystems at the same time [18]. They help in continual updating of entrepreneurial processes. Our findings concur with the literature that digital platforms create fluid boundaries of the entire entrepreneurial process as the products and services keep getting modified even after they have been introduced in the market and to the end user. This is done through modification in digital artifacts or components [24].

The limited resources are an important challenge that almost all small startups face. They have the required skillset, social network and creativity but are unable to have access to costly new resources [10]. Having enough finances to turn a viable prototype into a tangible product is a challenge for which help from pre-seed rounds is needed. However, various incubation and accelerator programs help overcome the financial challenges faced by digital platform-based startups. Combining available resources (internal and external) and seeking help from the limited social network can help in operationalization, improving sales performance and efficiency and creating customers. Technological challenges are also common in Pakistan where technology is still in its nascent stage. It is a little challenging to enhance platform user base in Pakistan as customers are at times less familiar with the technology either because it is expensive or due to less adequate technological knowledge. Making the technology comparative in terms of price point and entry and allowing the customers to pay only for what they use can help platform-based startups in overcoming technological resource challenges throughout the entrepreneurial process.

Startups in developing countries like Pakistan are more inclined towards operating on digital platforms as it enables entrepreneurship even in resource constraint environment. Scarce resources are a huge challenge and people in developing countries tend to have technology acceptance issues. However, combining available resources in an effective manner and taking help from existing network of family and friends can help in operationalizing the startups. Moreover, digital platforms charging customers only for what they use becomes a major solution to cater to the technology acceptance issues and enhancing user base.

In this article, we try to bring the attention of IS scholars towards exploring the interaction of digital platforms and entrepreneurial processes. It is a right time to focus on digital entrepreneurship as the entire world is moving towards digitization. In developing countries, startups and small enterprises can catalyst the economic and social development. This initial level research identifies the need of future research to unpack the various stages of entrepreneurial process from input, market and institutional resources facilitated by digital platforms in developed and developing countries. These theoretical and practical insights will not only contribute into emerging literature of digital entrepreneurship but also assist in flourishing the entrepreneurial ecosystems in developing countries where resources are scarce.